\documentclass[12pt]{article}

\usepackage{amssymb,amsmath,mathrsfs,epstopdf,slashed,color}

\usepackage{graphicx}
\usepackage{hyperref}
\usepackage{cite}

\allowdisplaybreaks[1]

\makeatletter

\@addtoreset{equation}{section}
\makeatother

\def\a{\alpha}

\title{\bf A Global View on The Search of de-Sitter Vacua in Type IIA String Theory}
\author{}

\begin{document}

\begin{titlepage}

\setcounter{page}{0}

\begin{flushright}
 \small
 MAD-TH-11-10\\
 \normalsize
\end{flushright}

\vskip 3cm
\begin{center}

  {\Large \bf A Global View on The Search for de-Sitter Vacua\\in (Type IIA) String Theory}

\vskip 2cm

{\large Xingang Chen${}^1$, Gary Shiu${}^{2,3}$, Yoske Sumitomo${}^3$, and S.-H. Henry Tye${}^{3,4}$}

 \vskip 0.7cm

 ${}^1$ Center for Theoretical Cosmology, Department of Applied Mathematics and Theoretical Physics, University of Cambridge, Cambridge, CB3 0WA, UK\\
 ${}^2$ Department of Physics, University of Wisconsin, Madison, WI 53706, USA\\
 ${}^3$ Institute for Advanced Study, Hong Kong University of Science and Technology, Hong Kong\\
 ${}^4$ Laboratory for Elementary-Particle Physics, Cornell University, Ithaca, NY 14853, USA

 \vskip 0.8cm


\abstract{\normalsize
The search for classically stable Type IIA de-Sitter vacua typically starts with an ansatz that gives  Anti-de-Sitter supersymmetric vacua and then raises the cosmological constant by modifying the compactification. As one raises the cosmological constant, the couplings typically destabilize the classically stable vacuum, so the probability that this approach will lead to a classically stable de-Sitter vacuum is Gaussianly suppressed. This suggests that classically stable de-Sitter vacua in string theory (at least in the Type IIA region), especially those with relatively high cosmological constants, are very rare. The probability that a typical de-Sitter extremum is classically stable (i.e., tachyon-free) is argued to be Gaussianly suppressed as a function of the number of moduli.

  }

\vspace{3cm}

\begin{flushleft}
 \today
\end{flushleft}

\end{center}
\end{titlepage}

\setcounter{page}{1}
\setcounter{footnote}{0}

\tableofcontents

\parskip=5pt

\section{Introduction}

Recent cosmological data strongly suggests that our universe is sitting at a vacuum state with a very small positive vacuum energy density, or cosmological constant. Further cosmological data also suggests that our universe went through an inflationary epoch in its very early stage. This epoch follows from the presence of a vacuum energy density much bigger than today's value. So it is very likely that our universe started with a relatively large vacuum energy density during the inflationary epoch; it subsequently moved down a ``potential landscape" and landed at the present small value before nucleosynthesis time. A question of great interest is whether this picture
is compatible with our present understanding of string theory. In particular, we would like to examine whether the stringy cosmic landscape has features that may suggest the above cosmological scenario.

Although recent studies of flux compactification in string theory suggest that there are numerous solutions to the string/supergravity equations of motion with different vacuum energies \cite{Bousso:2000xa,Kachru:2003aw,Susskind:2003kw}, we believe that most of them are only extrema of the resulting effective potential. In fact, explicit model building shows that a meta-stable (i.e., classically stable but may have a finite decay time due to quantum tunneling) de-Sitter ($dS$) vacuum is hard to come by. In particular, the search of a single $dS$ minimum (i.e., with only semi-positive scalar field mass-squares) in Type IIA models has so far come up empty. This is somewhat discouraging as the search includes a collection of exponentially many extrema (by varying the fluxes) in Type IIA vacua. On the other hand, this result is really not that surprising from the properties of multidimensional potentials \cite{Aazami:2005jf,Easther:2005zr}.

Consider a flux compactification with $N$ moduli. An extremum will be (meta-)stable if all the scalar mass-squares are semi-positive. The axionic component of each modulus presumably has an oscillating behavior, so it will hit a minimum half the time. If we hit a maximum, we expect a nearby minimum to which the wavefunction of the universe can easily move. To simplify the discussion, we may assume that it is easy to reach an axionic minimum and we focus mostly only on the real moduli. Now (with canonical kinetic terms), a typical stringy effective potential $V(\phi_j)$ ($j=1,2, \cdots ,N$) of $N$ (real) moduli have non-trivial behavior. Since none (or almost none) of the moduli takes a constant value, they are expected to have some extrema too. For a complicated potential $V(\phi_j)$ sitting at a minimum, the Hessian (i.e., the $N \times N$ mass-squared (symmetric) matrix, or simply mass matrix) must have only semi-positive eigenvalues. However, this likelihood is very small, as first pointed out by \cite{Aazami:2005jf}. Let ${\cal P}$ be the probability that a given de-Sitter solution (an extremum of a positive $V(\phi_j)$) turns out to be a meta-stable $dS$ minimum (that is, the $dS$ vacuum is tachyon-free).
To avoid de-compactification, we consider only the meta-stable vacua within the finite ranges of the moduli.
Suppose all the real entries in the Hessian is random, then the probability ${\cal P}$ that it has only positive eigenvalues is roughly given by
\begin{equation}
\label{intro1}
  {\cal P}  \sim   e^{- \frac{\ln 3}{4} (N+ 0.7)^2 }
\end{equation}
where $\ln (3)/4= 0.275$ is obtained in \cite{Dean:2006wk, Dean2008}. For a relatively large $N$, ${\cal P}$ is Gaussian-suppressed. Even if the moduli do not couple to each other so the Hessian is diagonal, the probability ${\cal P}$ for large $N$ is still exponentially suppressed,
\begin{equation}
\label{intro2}
  {\cal P}  \sim   \left( \frac{1}{2} \right)^N = e^{- N {\ln 2}}
\end{equation}
where $\ln 2= 0.693$. These probabilities apply to searches via trial and errors only.

One may argue that a generic potential $V(\phi_j)$ must hit some minima somewhere. This is certainly true. In the Type IIA cases, we see that all known (meta-stable) minima happen to have zero (Minkowski) or negative (Anti-de Sitter) vacuum energy densities, where supersymmetry as well as other symmetries help to guide the search for minima. In fact, some searches start from an $AdS$ minimum with only positive mass squares and then lift it to de-Sitter space. However, upon lifting to de-Sitter space, tachyon generically appears. The couplings among the moduli introduces off-diagonal terms in the Hessian. As the cosmological constant increases, the magnitudes of the off-diagonal terms increase as well, and that tend to cause instability.

To see the impact on the stability of the vacuum due to the increase of the vacuum energy, let us start with the diagonal positive mass-square matrix $A$ for an $AdS$ vacuum. Given the masses, we can determine the variance $\sigma_A$ of $A$. As we lift the cosmological constant, the Hessian at the extremum becomes $A+B$ where $B$ may be treated as a random matrix for a complicated generic $V(\phi_j)$. The matrix $B$ has variance $\sigma_B$. This allows us to define the size of the average magnitude of the off-diagonal terms relative to the diagonal mass-squared terms in terms of $y= \sigma_B/\sigma_A$.
The parameter $y$ essentially describes how the uplifting potential impact on the stability of the vacuum.
Now let
\begin{equation}
{\cal P} = a\, e^{-b N^2 - c N}
\end{equation}
where the Gaussian suppression dominates the exponential suppression when $bN/c >1$.
Numerically we find that, for small $y$,
\begin{equation}
 \begin{split}
    b&= 0.000395 y + 1.05 y^2 - 2.39 y^3,\\
  {b\over c}&= 0.0120+ 2.99 y - 12.2 y^2 + 1650 y^3.
 \end{split}
  \label{p suppressed off-diagonal comp}
\end{equation}
Here we see that Gaussian suppression becomes dominant when $y \ge 0.0241$ for $N=10$ (where $b/c \ge 0.1$ and $Nb/c \ge 1$) and $y \ge 0.00269$ for $N=50$. For fixed $N$, Gaussian suppression becomes more dominant as $y$ increases.

As an example, we look at a concrete search for a $dS$ vacuum starting from an $AdS$ vacuum undertaken in \cite{Caviezel:2008tf}. In this $SU(2) \times SU(2)$ model, there are 14 moduli (7 complex moduli), i.e., $N=14$. At the extremum with positive vacuum energy, $y \simeq 0.274$, so $bN/c \gg 1$.
This indicates that such a search has a Gaussianly small probability of success. Sure enough, tachyon appears at this extremum.

On the other hand, some generic argument suggests that $dS$ vacua  exist in Type IIB models, especially when non-perturbative effects are turned on.
For example, a KKLT vacuum may be obtained by uplifting an $AdS$ minimum with  non-perturbative effects. 
Attempts to construct $dS$ vacua in Type IIA models studied so far do not include non-perturbative effects.
This is because the no-scale structure is present at tree level in Type IIB, while not in Type IIA.
The no-scale structure may help to have a hierarchical structure in Type IIB with sub-leading corrections.
We shall discuss Type IIB models in a separate paper. Even without going into the details here, we shall use the simple belief that there are metastable $dS$ vacua present in string theory, even though they may be very rare.

Suppose all potentials of $dS$ vacua may be treated as ``uplifts" of $AdS$ vacua with semi-positive mass-squares. A small uplifting will introduce relatively small off-diagonal terms into the Hessian so the chance of being a $dS$ minimum is relatively good. As we increase the uplifting to higher vacuum energy densities, the off-diagonal terms in the Hessian increase accordingly and the Hessian becomes complicated. The off-diagonal terms tend to push the lowest eigenvalues to negative values. So the chance of this being a $dS$ minimum becomes Gaussianly small (i.e., (\ref{intro1})) as we go to higher cosmological constants. 
This leads to the conjecture that there are essentially no $dS$ minima in the relatively higher CC regions in the Type IIA cosmic landscape. The message leads to the following proposal :
\begin{quote}
 {\it Raising the cosmological constant destabilizes the classically stable vacua}.
\end{quote}
This suggests that, as the universe evolves down the potential, it encounters no $dS$ minima along its way for relatively large cosmological constant (CC) values. Presumably, this happened during the inflationary epoch. Towards the end of inflation, when the universe reaches regions with sufficiently small CC, there may be some $dS$ minima around for the universe to be trapped in one of them. That is, the percolation probability is of order unity for high CC, but decreases substantially by the time the universe reaches the small CC region.

It is very likely that there are many more $AdS$ vacua than $dS$ minima around, but since the universe starts from a relatively high CC region (to generate enough inflation), it has to go through the small positive CC region before reaching the negative CC region. It is not unreasonable that it becomes trapped in a low $dS$ minimum on its way towards negative CC region if the probability of finding a $dS$ minimum with a small CC is not too suppressed. As we shall see, it is reasonable to expect that, for the small positive CC region, some of the moduli with large masses essentially decouple, thus reducing $N$ to a smaller effective value, so the probability of finding a low $dS$ minimum may not be exponentially suppressed compared to the number of $AdS$ vacua nearby.

Since no classically stable Type IIA $dS$ vacuum has been found so far, the above proposal cannot be (non-trivially) checked at the moment. However, we do know some Type IIB solutions; so presumably such a check can be performed for Type IIB regions of the landscape. This study (which involves some subtleties) is under way.

This paper is organized as follows. In section \ref{sec:prob-estim-rand}, we review and discuss the properties of a real symmetric random matrix as a typical example of a Hessian. We then discuss a Hessian that is the sum of a random diagonal positive matrix plus a random symmetric matrix. This mimics the Hessian that generically appears in the search for a $dS$ minimum in Type IIA models. In Sec. 3, we consider a few examples to illustrate the main point of this paper.
Sec. 4 contains some discussions. In particular, we shall comment on an earlier estimate of
the probability of obtaining $dS$ vacua  in \cite{Denef:2004cf}. Some details are relegated to the appendix.

\section{Probability estimation in random matrix \label{sec:prob-estim-rand}}

We now discuss the probability ${\cal P} $ of a complete random mass matrix to have only positive eigenvalues, and then contrast the results with that for a mass matrix where the off-diagonal components between
 K\"ahler and complex or dilaton moduli are suppressed.
Note that the probability being considered is the probability that an extremum with positive vacuum energy density turns out to be a $dS$ minimum.

\subsection{A complete random matrix\label{sec:compl-rand-matr}}

When we cannot neglect off-diagonal components of a mass matrix, we expect tachyon directions to develop since the off-diagonal components
 repel
 the eigenvalues through
diagonalization.
Before studying more realistic examples and computing the mass matrix from concrete models, we first review how unlikely one can obtain minima in a complete random matrix \cite{Aazami:2005jf,Dean:2006wk, Dean2008}.

\begin{figure}[t]
 \begin{center}
  \includegraphics[width=18em]{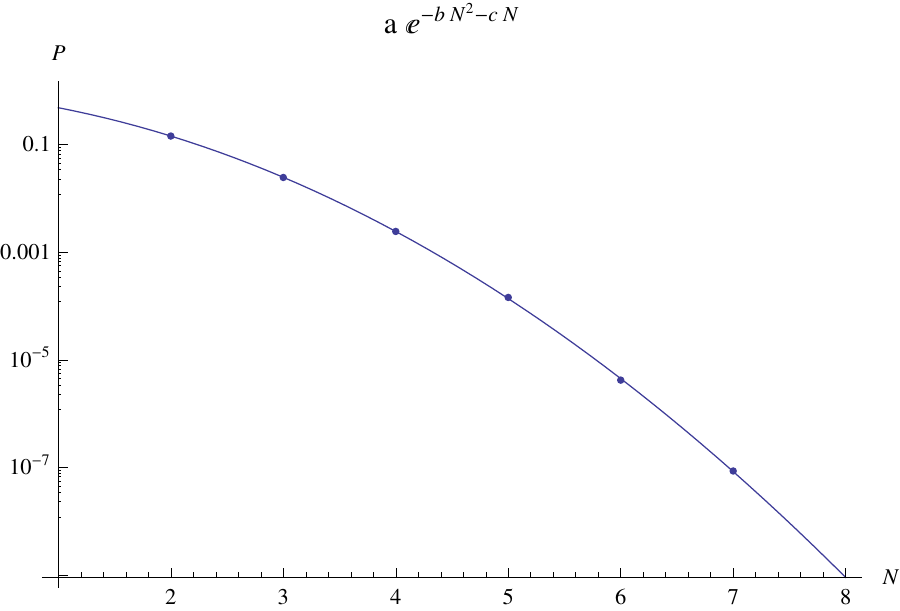}
 \end{center}
 \caption{\footnotesize The probabilities of positive definite real symmetric random matrices.
 Each diagonal entry is independently assigned a real number obeying the normal distribution with central value zero and variance one, while off-diagonal entries has variance $1/\sqrt{2}$, i.e., we assume a Gaussian orthogonal ensemble. The fitting function is $a e^{-b N^2 - c N}$ where $a \sim 0.938, b \sim 0.277, c\sim 0.382$.}
 \label{fig:randomprob}
\end{figure}

We are interested in counting the probability for a real-symmetric random matrix to be positive definite.
We assume that each diagonal component obeys the normal distribution with central value 0 and variance 1, while off-diagonal components have variance $1/\sqrt{2}$ ({\it Gaussian Orthogonal Ensemble} with variance 1).
We increase the size of the matrix and numerically count the probability of the positive definitive matrices as a function of the dimension of the matrix, $N$. The number $N$ represents the number of moduli in the realistic landscape.
Using the fitting function ${\cal P} = a e^{-b N^2 - c N}$, our simulation shows that, for $N\leq 7$,
\begin{equation}
 \begin{split}
  {\cal P} \equiv {{\rm \# \ of\ events} \over {\rm \#\ of\ trials}} \sim 0.938 \, e^{- 0.277 N^2 - 0.382 N} ~.
 \end{split}
 \label{probability of real-symmetric random matrix}
\end{equation}
See figure \ref{fig:randomprob}.

This fit reproduces well the theoretical prediction in \cite{Dean:2006wk} where $b = \ln (3) / 4 \sim 0.275 $ (for Gaussian orthogonal ensemble), although the prediction is applicable at large $N$.
In \cite{Dean:2006wk}, an eigenvalue fluctuation around the edge of Wigner semi-circle for finite $N$ (referred as Tracy-Widom \cite{Tracy:1992rf}) was taken into account, such that the density function of eigenvalues was corrected accordingly.
The analytic expression was applicable under the assumption of large $N$, i.e., for leading behavior proportional to $N^2$. However this is still a good estimation for small $N$ if we incorporate the linear dependence in the exponent as a correction for small $N$.
Interestingly, the complete form of analytical expression was achieved recently in \cite{Borot:2010tr}
\footnote{ 
We would like to thank David Marsh and Timm Wrase for bringing our attention to the paper.}.
For GOE, the probability function is given by
\begin{equation}
\begin{split}
{\cal P} =& \exp \left[ -{\ln 3 \over 4}N^2 + {\ln (2\sqrt{3}-3) \over 2} N - {1\over 24} 
\ln N - 0.0172  \right]\\ 
\sim& 0.983\, e^{-0.275 N^2 - 0.384 N - 0.0417 \ln N}.
\end{split}
\end{equation}
Therefore our numerical simulation mostly agrees with the analytical expression, not only for the coefficient of leading $N^2$ term, but also for the linear dependence and constant term, although we neglected $\ln N$ dependence which is actually small even between $N=2-7$.
For simplicity, if we fit a form ${\cal P} = e^{- {\ln 3 \over 4}(N+d)^2}$, then the choice of $d\simeq 0.7$ provides a good fit for ${\cal P}$ for all $N>1$.

\subsection{A random matrix with suppressed off-diagonal components\label{sec:random-matrix-with}}

As we start from an $AdS$ vacuum with only positive mass-squared eigenvalues and raise the vacuum energy in the search for a $dS$ vacuum, couplings among the moduli introduces terms into the Hessian, in particular off-diagonal terms.
We like to estimate ${\cal P}$ as a function of the relative size of the off-diagonal terms emerging in the Hessian.
In particular, we like to see when ${\cal P}$, as a function of $N$, is Gaussianly suppressed versus exponentially suppressed.
This leads us to consider a mass-squared matrix in which the magnitudes of diagonal components is larger than that of the off-diagonal components.
We again introduce randomness to mimic some of the features in the Hessian generically.

We consider the following mass-squared matrix:
\begin{equation}
 M = A+B
\end{equation}
where the matrix $A$ only has real diagonal components which obey half-normal distribution (positive definite) with variance $\sigma_A$, while each component of a real-symmetric matrix $B$ obeys Gaussian orthogonal ensemble with variance 1, same as before.
We use the half-normal distribution to model the statistical distribution of the diagonal elements at leading order. The difference between $\sigma_A$ and $\sigma_B=1$ introduces the hierarchy between the diagonal elements and off-diagonal elements.

The basic idea for this modeling is the following: first we consider moduli stabilization at $AdS$ minima by some mechanism, and then add an uplifting potential(s) to attain a positive cosmological constant.
The diagonal mass matrix $A$ is given at $AdS$ and the real-symmetric mass matrix $B$ comes from the uplifting term since the stabilized moduli masses are generically mixed in the presence of additional sources.
In the previous section, we consider situations in which
the mixing term is comparable to the diagonal terms in the potential.
Here, we consider scenarios where the mixing is suppressed, such that the diagonal entries are more likely to be positive.

We simulate the probabilities of positive definite mass matrix for $N=4-20$ while varying the variance of diagonal matrix between $\sigma_A = 10,15,20, \cdots, 100$. We choose the fitting function to be still of the form
\begin{equation}
{\cal P} = a\, e^{-b N^2 - c N} ~,
\label{P_fitting}
\end{equation}
but now the $b$ and $c$ are both functions of $\sigma_A$.
The motivation for this choice is as follows. For random matrices that we studied in the previous subsection, the Wigner semicircle law implies that the eigenvalues are mostly distributed within $[-2\sqrt{N},2\sqrt{N}]$. Now we have added some additional positive diagonal elements. If we fix the hierarchy between the diagonal and off-diagonal terms, but increase the dimension $N$, the above range will keep increasing and eventually swamp the fixed hierarchy we introduced. Namely we expect to recover the Gaussian suppression in the large $N$ limit. This is why we choose the leading term in the exponential to be still proportional to $N^2$. For smaller $N$, we expect it to be less Gaussian, and this is modeled by the linear term.

\begin{figure}[t]
 \begin{center}
  \includegraphics[width=18em]{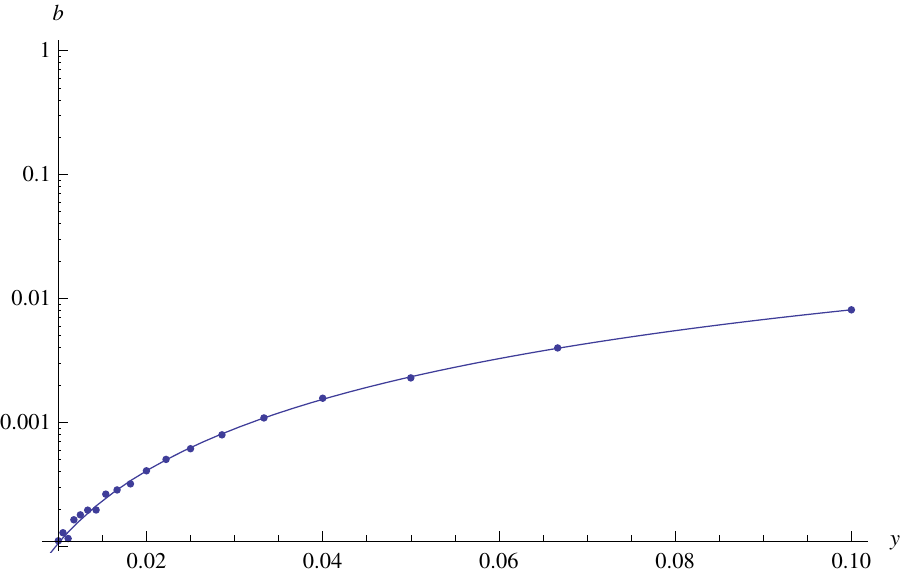}
  \hspace{1em}
  \includegraphics[width=18em]{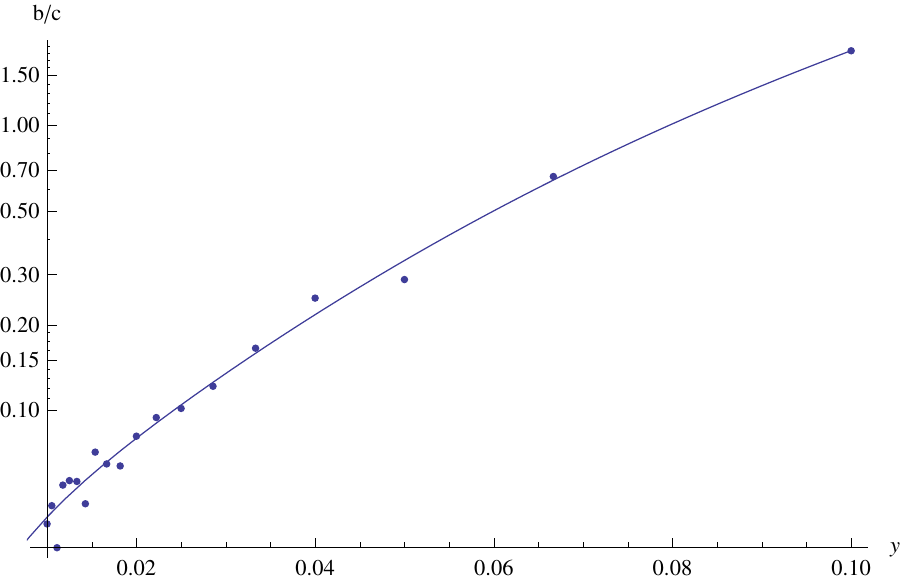}
 \end{center}
 \caption{\footnotesize The probability estimation for variances $\sigma_A = 10-100$ in the positive definite diagonal matrix $A$ with $N=4-20$.
 The variance of real-symmetric matrix $B$ is fixed at $\sigma_B = 1$.
 We use the probability fitting function of the form $P= a\, e^{-b N^2 - c N}$ as before, while each coefficient depends on the relative ratio of variances $y=\sigma_B/\sigma_A$.
 We found that $a$, $b$, $c$ can be fitted by (\ref{probability of suppressed off-diagonal comp})
   and $a$ is mostly one between $\sigma_A=10-100$.}
 \label{fig:small-offdiag}
\end{figure}

For example, for $\sigma_A=10$, we get
\begin{equation}
{\cal P} = 0.950\, e^{-0.00810 N^2-0.00442 N} ~;
\label{fit_example_1}
\end{equation}
for $\sigma_A=100$, we get
\begin{equation}
{\cal P} = 1.00\, e^{- 0.000111 N^2-0.00277 N} ~.
\label{fit_example_2}
\end{equation}
The coefficients $b$ and $c$ as functions of $\sigma_A$ (or equivalently $y\equiv \sigma_B/\sigma_A$) can be fitted by the following formulae, (see figure \ref{fig:small-offdiag}),
\begin{equation}
 \begin{split}
  b&= 0.000395 y + 1.05 y^2 - 2.39 y^3,\\
  {b\over c}&= 0.0120+ 2.99 y - 12.2 y^2 + 1650 y^3.
 \end{split}
  \label{probability of suppressed off-diagonal comp}
\end{equation}
The extrapolation of these fitting functions to larger values of $N$ works quite well. For example, for $N=30$, $y=1/10$, the expected probability from (\ref{fit_example_1}) is ${\cal P}_{\rm exp} = 0.00056$, while the numerical simulation suggests ${\cal P}_{\rm obs} \approx 0.00048$; for
$N=150,\ y=1/100$, the expected probability is ${\cal P}_{\rm exp}=0.0548$, while the observed probability is ${\cal P}_{\rm obs}= 0.0318$.
In the large $N$ limit with fixed $y$, we still have the same form (\ref{P_fitting}) but we expect the formulae for the coefficients $b$ and $c$ (\ref{probability of suppressed off-diagonal comp}) to deform and approach the limit of the complete random matrix.
These fitting values may also be applicable even if we decrease $y$ value further.
At $y=1/200$ and $N=300$ which is outside of the data used for the fitting (\ref{probability of suppressed off-diagonal comp}), the expected probability goes as ${\cal P}_{\rm exp}=0.0600$.
On the other hand, the observed probability is given by ${\cal P}_{\rm obs}=0.0202$, therefore not so bad.

We conclude that the probability decreases immediately once the off-diagonal components become non-negligible.
From the data in figure \ref{fig:small-offdiag}, the ratio of coefficients is reaching to $b/c=0.1$ at $y=0.0241$.
This is the point when the $N^2$ term is comparable to the linear term at $N=10$.
Therefore we are very likely to encounter a tachyon when off-diagonal components are not suppressed with $y \le 0.0241$.
If a model of interest has a lot more moduli, for instance $N=50$, then the Gaussian suppression in the probability is expected unless the off-diagonal suppression becomes less than $y=0.00269$.
On the other hand, when the coefficient $b$ is comparable to the value of $c$, the probability is mostly Gaussianly suppressed even at lower $N$.

 Another possible fitting function for the probability ${\cal P}$ is given in
 appendix \ref{sec:other-plot-random}, which however does not have the expected large $N$ behavior and thus work not well when extrapolating to large $N$.

In the next section, we consider some concrete models in IIA.
In these models, the mass matrix at the extrema
is well mimicked by a
 random matrix  analysis.

\section{Search of Type IIA de Sitter vacua}

There have been substantial efforts in constructing de Sitter vacua from Type IIA string theory.
 The simplicity of Type IIA flux vacua makes them more amendable to detailed, quantitative analysis.
In contrast to Type IIB compactifications on warped Calabi-Yau spaces,
classical effects  (i.e,, from fluxes) alone can stabilize  both the complex
 structure and Kahler moduli in Type IIA string theory.
Motivated by the infinite family of $AdS$ Type IIA flux vacua found in \cite{DeWolfe:2005uu},
attempts to construct de Sitter flux vacua were made in \cite{Silverstein:2007ac,Haque:2008jz,Caviezel:2008tf,Flauger:2008ad,Danielsson:2009ff,Caviezel:2009tu,deCarlos:2009fq,deCarlos:2009qm,Dong:2010pm,Andriot:2010ju,Danielsson:2010bc,Danielsson:2011au}.
We refer to such de Sitter vacua which do not invoke non-perturbative effects as {\it classical de Sitter vacua}.
The absence of non-perturbative effects and explicit supersymmetry breaking localized sources\footnote{There has been some recent progress in finding backreacted solutions of anti-3 branes
 \cite{DeWolfe:2008zy,McGuirk:2009xx,Bena:2009xk,Bena:2010ze,Dymarsky:2011pm,Bena:2011hz,Bena:2011wh} (see also the discussion about anti-6 brane \cite{Blaback:2011nz,Blaback:2011pn}) and 10D lift of instanton effects \cite{Koerber:2007xk,Heidenreich:2010ad,Dymarsky:2010mf}.} makes it possible to
 construct and describe such classical de Sitter vacua as
10D backgrounds, rather than candidate minima of an effective 4D potential.
Moreover, the backreaction of fluxes on the background can be
captured by the framework of generalized complex geometry,
allowing us to make use of the strong mathematical results previously obtained on the subject.

The idea is to consider Type IIA string compactifications on internal spaces with negative  curvature.
Various no-go theorems for de Sitter extrema have appeared in the literature \cite{Gibbons:1984kp,Maldacena:2000mw,Hertzberg:2007wc}.
By revisiting the assumptions made, it was found that the minimal ingredients needed to
evade such no-goes are fluxes, orientifold planes, and a negatively curved internal
space  \cite{Haque:2008jz,Danielsson:2009ff,Wrase:2010ew}.
Interestingly, these minimal ingredients can be realized in compactificatons on generalized complex geometries.
The metric fluxes which distinguishes these generalized geometries from Calabi-Yau spaces
 also induce  curvature in the internal space. Many of these generalized geometries are negatively curved, at least in some region of the moduli space.
Though less familiar than Calabi-Yau spaces,
generalized complex geometries are more general supersymmetric backgrounds for string theory
(e.g., orientifold compactifications on SU(3) structure manifolds leads to an effective ${\cal N}=1$ SUGRA, just as Calabi-Yau orientifolds
\footnote{The effective potential was obtained for IIA Calabi-Yau orientifolds \cite{Grimm:2004ua} and for non-Calabi-Yau orientifolds \cite{Benmachiche:2006df}.}).
Of particular interest are spontaneous supersymmetry breaking states of such effective 4D supergravities as they are amendable to
the powerful tools of supersymmetry. Note also that the no-go theorems in
\cite{Hertzberg:2007wc,Haque:2008jz,Danielsson:2009ff,Wrase:2010ew} concern only with the existence of de Sitter critical points. It was shown in \cite{Shiu:2011zt} that the stability of such candidate de Sitter vacua provide further strong constraints on the choice of flux vacua.

While we emphasized above the importance of being able to lift
a 4D de Sitter construction to a full solution of the 10D equations of motion,
we review the 4D description of classical de Sitter solutions here as it greatly facilitates a comparison with our Type IIB analysis presented in a forthcoming paper. 4D SUSY implies the existence of a globally well-defined internal 6D spinor, though such spinor is not necessarily convariantly constant. Thus, the real 2-form $J$ and the complex 3-form $\Omega=\Omega_R + i \Omega_I$ can be constructed from such a spinor; they define an $SU(3)$ structure (and not an
$SU(3)$ holonomy in general) and   are not necessarily closed:
\begin{equation}
\begin{split}
d J =& - \frac{3}{2} {\rm Im} (\overline{W}_1 \Omega) + W_4 \wedge J + W_3, \\
d \Omega =& W_1 J \wedge J  + W_2 \wedge J  + \overline{W}_5  \wedge \Omega.
\end{split}
\end{equation}
The torsion classes $W_1, \dots, W_5$ correspond to the expansion  of the derivatives of $J$ and $\Omega$ in terms of $SU(3)$ representations: $W_1$ is a complex scalar, $W_2$ is a complex primitive $(1,1)$ form, $W_3$ is a real primitive $(1,2) + (2,1)$ form, $W_4$ is a real one-form, and $W_5$ is a complex $(1,0)$ form. An $SU(3)$ structure (that is not also an $SU(2)$ structure)
manifold has no nowhere-vanishing one-forms. If we restrict to
such torsion classes,
$W_4$ and $W_5$ must vanish.

When the internal space is a non-Ricci flat $SU(3)$ structure manifold, there are some subtleties with the identification of the light fields and the associated low-energy effective action \cite{KashaniPoor:2006si,KashaniPoor:2007tr}.
However, for group manifolds (and coset spaces) which we will focus on, we can restrict to expansion forms that are left-invariant under the group action. This leads to a 4D theory that is a consistent truncation \cite{Cassani:2009ck}, i.e., a solution to the 4D equations of motion will also be a solution to the 10D equations of motion.
Under the orientifold symmetry,
$\Omega \rightarrow - \Omega^*$, and $J \rightarrow -J$.
Hence, we can expand $J$ and $\Omega$ in terms of a representative basis of forms:
\begin{equation}
\begin{split}
J =& k^i Y_i^{(2-)}, \\
\Omega =& {\cal F}_K Y^{(3-)}_K + i {\cal Z}^K Y^{(3+)}_K,
\end{split}
\end{equation}
where $Y_i^{(2 \pm)}$, $i=1,\dots,h^{(1,1)}_-$ is a set of two-forms even (odd) under the orientifold parity,
and $Y_K^{(3 \pm)}$, $K=1,\dots, h^{(2,1)}+1$ is a set of three forms which are even (odd) under orientifolding.
Note that ${\cal F}_K$ are functions of the ${\cal Z}_K$ and therefore not independent.

The effective 4D SUGRA resulting from reduction on an $SU(3)$-structure space is completely specified by the superpotential W, the Kahler potential K,
a set of gauge kinetic functions $f_{\alpha, \beta}$, and their associated  D-terms $D_{\alpha}$.
\begin{equation}
\begin{split}
K =&  -2 \, \ln  \left( -i \int e^{-2\phi} \Omega \wedge \Omega^* \right) -\ln \left( \frac{4}{3} \int J \wedge J \wedge J \right) = 4 \phi_4 - \ln \left( 8 \text{vol}_6\right),\;\;\\
\sqrt{2} W = & \int \left( \Omega_c \wedge (-i H + d J_c) + e^{i J_c} \wedge \hat{F} \right),\\
f_{\alpha \beta} =& - \hat{\kappa}_{i \alpha \beta} t^i,\\
D_\a =& - \frac{e^{\phi_4}}{\sqrt{2 \text{vol}_6}} \hat{r}_{\alpha}^K {\cal F}_K,
\end{split}
\end{equation}
where $\phi_4$ is the 4D dilaton defined by $e^{-\phi_4}=
e^{-\phi} \sqrt{\text{vol}_6}$ with $\Omega \wedge \Omega^* = (4i/3) J  \wedge J \wedge J=8 i \text{vol}_6$, and $\hat{F}=\hat{F}_0 + \hat{F}_2 +
\hat{F}_4 + \hat{F}_6$ is the sum of the RR fluxes.
The 2-form $J_c$ and 3-form $\Omega_c$ that appear in the superpotential above are given by combinations with other supergravity fields, namely, the dilaton $\phi$, the Kalb-Ramond two-form $B$, and the RR three-form $C_3$:
\begin{equation}
\begin{split}
J_c =& J - i B = t^i Y_i^{(2-)}, \\
\Omega_c =& e^{-\phi} {\rm Im} (\Omega) + i C_3 = N^K Y_K^{(3+)}.
\end{split}
\end{equation}
The triple intersection numbers which enter into the gauge kinetic function are defined in terms of the basis forms:
\begin{equation}
\kappa_{ijk} = \int Y^{(2-)}_i \wedge Y^{(2-)}_j \wedge Y^{(2-)}_k, \qquad \hat{\kappa}_{i \alpha \beta} = \int Y^{(2-)}_i \wedge Y^{(2+)}_{\alpha} \wedge Y^{(2+)}_{\beta}.
\end{equation}
The D-terms
contain information about the metric fluxes.
The matrices $r_{iK}$ and $\hat{r}_{\alpha}^K$ are
 defined as follows:
\begin{equation}
d Y^{(2-)}_i = r_{iK} Y^{(3-)K},\qquad d Y^{(2+)}_{\alpha} = \hat{r}_{\alpha}{}^K Y^{(3+)}_K.
\end{equation}
On an $SU(3)$ structure group/coset manifold (the type of $SU(3)$ structure manifolds where explicit examples have been constructed), there exist
 six global left-invariant one-forms $e^a$, $a=1,\dots,6$ and the metric fluxes $f^a_{bc}$ are introduced through
 $d e^a = - \frac{1}{2} f^a_{bc} e^b \wedge e^c$. The matrices
 $r_{iK}$ and $\hat{r}_{\alpha}^K$
 are therefore linear functions of the metric fluxes.

The 4D scalar potential for the left-invariant modes
then follows from the usual supergravity expression:
\begin{equation}
V = e^K \left( K^{ij} D_{t^i} W \overline{D_{t^j} W} +K^{K\overline{L}}  D_{N^K} W \overline{D_{N^L} W} - 3 |W|^2 \right) + \frac{1}{2} \left( {\rm Re} f \right)^{-1^{\alpha \beta}} D_{\alpha} D_{\beta}
\label{SUGRAV}
\end{equation}
 where the derivatives $D_{t^i} W = \partial_{t^i} W + W \partial_{t^i}
K$ (and analogously for $D_{N^K}$).

Even though the 4D effective action of $SU(3)$ structure compactifications may appear to be
more complicated in form than their Calabi-Yau counterpart, each individual term in the action can be explicitly computed given the geometric and flux data.
This is an advantage over the more well studied Type IIB scenarios where non-perturbative instanton corrections and the effects of SUSY breaking localized sources are
often not computed in explicit detail.
In particular, useful results can
be readily obtained by analyzing how various contributions to the 4D potential scale with the moduli.
Most of such analysis was carried out for the universal moduli subspace as these moduli appear in any Type II compactification. Consider the following metric ansatz in the 10D string frame:
\begin{equation}
ds_{10}^2 = \tau^{-2} ds_4^2 + \rho d s_6^2
\end{equation}
where we took the Weyl factor to be $\tau = e^{-\phi} \rho^{3/2}$ such that the kinetic terms for the universal moduli $\rho$ and $\tau$ in the 4D Einstein frame do not mix.

Various fluxes $H_3$, $F_q$, localized $q$-brane sources and the 6D curvature contribute to the 4D potential in some specific way:
\begin{equation}
  V_{H_3} = A_{H_3} \tau^{-2} \rho^{-3}, \quad V_{F_p} = A_{F_p} \tau^{-4} \rho^{3-p}, \quad V_q = A_q \tau^{-3} \rho^{(q-6)/2}, \quad V_{R_6} = A_{R_6} \tau^{-2} \rho^{-1}.
   \label{potentials in two moduli}
 \end{equation}
The coefficients $A_{H_3}$ and $A_{F_q}$ of the flux potentials are defined to be positive, while the coefficients $A_q$ and $A_{R_6}$
can be either positive or negative.
 Note that all the potentials go to zero when taking $\tau\rightarrow \infty$ while keeping the others finite.
 Therefore there are always Minkowski vacua asymptotically.
From these scalings, we can already derive some simple no-go theorems for de Sitter vacua by finding a differential operator $D \equiv -a \tau \partial_{\tau} - b \rho \partial_{\rho}$ such that $D V \geq c V$ for some non-trivial real constants $a$, $b$, and $c>0$. However, upon closer inspection, one finds, using the Sylvester's criterion on the universal moduli subspace, that the minimal ingredients which evade the no-goes only guarantee that de Sitter extrema are allowed.
Additional ingredients are needed to warrant stability. The diagnostic advocated in \cite{Shiu:2011zt} allows one to rule out a lot of models at the outset without going into the details of the constructions
\footnote{The same method was applied for higher dimensional $dS$ vacua \cite{VanRiet:2011yc}.}.

To extend the stability analysis of \cite{Shiu:2011zt} to more moduli, we need to specify a model.
For reasons above (e.g., to ensure that the $SU(3)$ structure reduction is a consistent truncation), and in order to construct the $SU(3)$ structure explicitly, the searches for de Sitter have so far been
limited to homogeneous spaces (e.g., group manifolds, and coset spaces).
A classification of $SU(3)$ structure group manifolds was carried out in \cite{Danielsson:2011au}.
The number of 6D unipotent\footnote{A Lie algebra is unipotent if the structure constants are traceless, i.e., $f^a_{ab}=0$. Unipotence is a necessary condition for the group G to admit a freely-acting discrete subgroup L, such that $G/L$ is compact.}
group spaces
considered in \cite{Danielsson:2011au} is of the order of 50.
Quotienting such group spaces by orbifolds which preserve ${\cal N}=1$ supersymmetry
and which evade the no-goes for de Sitter extrema puts additional constraints on the structure constants and only 5-10 of the 6D nilpotent group spaces remain as viable candidates.
Taking as a rough estimate that each of these group spaces has ${\cal O} (10)$ left-invariant modes, and that tadpole conditions require that the flux quanta supported on each cycle to be  $\leq {\cal O}(10)$, the sample space of $SU(3)$ structure being studied is of the order  $10^{10}$.
In reality, a much smaller number of such solutions have been (and need to be) explicitly constructed as the constraints\footnote{Motivated by the form of some known SUSY $AdS$ solutions, a universal ansatz was taken to facilitate the search for de Sitter solutions. The constraint equations imply relations between contributions to the equations of motion and allow for new solutions other than the SUSY ones.} on the torsion classes and fluxes which allow for new solutions other than the known SUSY $AdS$ solutions already rule out many candidate de Sitter solutions before flux quantization and tadpole conditions are imposed.

Earlier studies (through explicit models, or general arguments using Sylvester's criterion on the universal moduli subspace)
showed that tachyons are ubiquitous in classical de Sitter solutions.
The main reason is the relatively unsuppressed off-diagonal terms in the moduli mass matrix.
Small numbers can be generated with $\alpha'$ and non-perturbative corrections.
In order for the classical SUGRA approximation to be valid, these corrections have to be sufficiently small (i.e., large volume, weak coupling, etc).
If there exist symmetries that forbid the off-diagonal terms to appear at the classical level, they are naturally suppressed (by the smallest of $\alpha'$ and non-perturbative corrections).
This so far has not happened (at least for the subset of moduli we analyzed with the Sylvester's criterion) as off-diagonal elements are present already classically.
(Though there may be classical de Sitter solutions
where the off-diagonal terms are absent at leading order. It would be interesting to make this a requirement in the search).

  \subsection{Mass matrices in concrete models}

  We focus on some concrete models to compare with our random matrix analysis in detail.

   \subsubsection{$SU(2)\times SU(2)$ group manifold on orientifold}

   In this subsection, we focus on an $SU(2)\times SU(2)$ group manifold described in \cite{Caviezel:2008tf}.
   This model
   evades
   the no-go theorem for extrema (see e.g. \cite{Hertzberg:2007wc,Wrase:2010ew})
   and thus was considered a candidate for metastable $dS$ vacua.
   However
   only tachyonic solutions were found
   in the search
   for $dS$ extrema
    performed in \cite{Caviezel:2008tf}.
   Now we will analyze the mass matrix of this model at an extremal point, and see
   how it depends on the ratio of variances
   between the off-diagonal entries and the diagonal ones.

   Defining two-forms $Y_i^{(2-)}$ and three-forms $Y^{(3-)I}$ on orientifolds of $SU(3)$-structure manifolds, where $i=1,2,3$ while $I=1,2,3,4$.
   Owing to the $SU(2)\times SU(2)$ symmetry of the manifold, which is our focus here, the two-forms and the three-forms satisfy the relation:
   \begin{equation}
    \begin{split}
     dY_i^{(2-)} = r_{iI} Y^{(3-)I}, \quad
     r=\left(\begin{array}{cccc}
        1&1&1 &-1 \\1 & -1&-1&-1\\1&-1&1&1
             \end{array}\right).
    \end{split}
   \end{equation}
   The background fluxes are
   given by
   \begin{equation}
    \begin{split}
     &F_0=m, \quad F_2 = m^i Y_i^{(2-)}, \quad F_4=0, \quad F_6=0,\\
     &H=p \left(Y_1^{(3-)}+Y_2^{(3-)}-Y_3^{(3-)}+Y_4^{(3-)}\right).
    \end{split}
   \end{equation}
   Using these inputs, the K\"ahler potential and superpotential become
   \begin{equation}
    \begin{split}
     K=&-\ln \prod_{i=1}^3 (t^i+\bar{t}^i) - \ln \prod_{I=1}^4 (N^I+\bar{N}^I) + 3\ln \left({ \kappa_{10}^2 \over V_s} \right) + \ln 32 ,\\
     W=& {V_s \over 4 \kappa_{10}^2} \left[ m^1 t^2 t^3 + m^2 t^1 t^3 + m^3 t^1 t^2 - i m t^1 t^2 t^3 - i p (N^1 + N^2 - N^3 + N^4) + r_{iI}t^i N^I\right],
    \end{split}
   \end{equation}
   where $V_s = \int_M Y^{(2-)}_1 \wedge Y^{(2-)}_2 \wedge Y^{(2-)}_3$.
   The potential can be calculated by
   \begin{equation}
    V = e^{K} \left(K^{A\bar{B}} D_A W D_{\bar{B}} \bar{W} - 3 |W^2|\right)
   \end{equation}
   without the
   D-terms.

   In this 14 moduli system which consists of $t^i= k^i- i b^i$ and $N^I = u^I + i c^I$, a positive extremum was found by \cite{Caviezel:2008tf}
   \begin{equation}
    \begin{split}
     &m^1 = m^2 = m^3 = L, \quad m = 2 L^{-1}, \quad p = 3 L^2,\\
     &k^1= k^2 = k^3 \sim 0.8974 L^2, \quad b^1 = b^2 = b^3 \sim - 0.8167 L^2,\\
     &u^1 \sim 2.496 L^3, \quad u^2 = -u^3 = u^4 \sim -0.5667 L^3,\\
     &c^1 \sim -2.574 L^3, \quad c^2 = -c^3 = c^4 \sim 0.3935 L^3,
    \end{split}
   \end{equation}
   where $L$ is a parameter assigned for the solution.
   The solution contains one tachyonic direction which shows up after diagonalizing the mass matrix.
   It turns out that all $2 \times 2$ and $3 \times 3$ sub-matrices have positive determinants.
   The tachyon first appears in the $4\times 4$ sub-matrix for the real parts of the complex moduli.

   Let us compare the mass matrix of this model with the random matrix considered in section \ref{sec:random-matrix-with}.
   Changing the basis of the mass matrix to the canonical one: $X^{i, I}, Y^{i, I}$ through the relations $dX^i = {1/2k^i} (d k^i + d b^i), \ dX^I = {1/2u^I}(du^I + dc^I), \ dY^i =  {1/2k^i} (d k^i - d b^i), \ dY^I = {1/2u^I}(du^I - dc^I)$ at the extrema, we see that only the overall factor of the mass matrix depends on the parameter $L$.
   After calculating the deviations of the mass matrix $M_{AB}$ assuming the center value to be zero, we get
   \footnote{The relative ratio calculated here is slightly different from the ratio $y$ considered in section \ref{sec:random-matrix-with}.
   This is because the uplifting matrix $B$ also includes diagonal entries, therefore deviation of diagonal components in the total matrix is not the one for $A$.
   However, since the deviation of $B$ is smaller than that of $A$, this is not a bad approximation.}
   \begin{equation}
    y \sim \left( {{1\over 14 \times 13/2} \sum_{A < B} M_{AB}^2 \over {1\over 14} \sum_{A=1}^{14} M_{AA}^2 }\right)^{1/2} = 0.274.
   \end{equation}
   The numerical value of the relative ratio $b/c$ obtained from
    (\ref{probability of suppressed off-diagonal comp})
    may not be applicable here since $y=0.274$ is already outside the domain of our random matrix estimation for small $y$.
   However, we know that the probability ${\cal P}$ is Gaussianly suppressed already at $y=0.1$ in (\ref{probability of suppressed off-diagonal comp}), where $b/c \simeq 1.8$. So we expect that, for $y =0.274$, ${\cal P}$ is Gaussianly suppressed even around small $N>1$.
   Thus it is not so surprising to have a tachyon at $N=14$ in this model.

Let us make a couple of comments here :
\begin{itemize}
\item The axionic directions with oscillating type potentials typically will have many minima.
This tends to provide stability along those directions. So one may exclude the number of axions in the effective $N$ used in the estimate of ${\cal P}$. In the above model with 4 complex moduli, we may include the real parts of the Kahler (and dilation) moduli and both the real and imaginary parts of the complex moduli: $N=3+2(4)=11$, instead of $N=14$.

\item The inflationary slow-roll parameter $\eta$ of the above model is given in \cite{Caviezel:2008tf}. For the potential $V>0$,
  \begin{equation}
  \eta = ({\rm min\ eigenvalue}) \times \frac{ \nabla^A \partial_A V}{V} \lesssim -2.4
   \end{equation}
Here, $\eta$ is negative, indicating the presence of a tachyon. The relatively slowly varying $\eta$ measures the ratio of the tachyon mass-squared with respect to the value of the positive vacuum energy. That is,  as we raise the vacuum energy of the potential $V$, the tachyon mass-squared become more negative correspondingly so the extremum becomes more unstable. This indicates that the increase of the vacuum energy tends to destabilize the vacuum.
\end{itemize}

   \subsubsection{A simple model with two moduli}

   Next we consider
   the two universal moduli subspace
    to clarify
    how the off-diagonal components increase as
    we increase the positive cosmological constant.
   The conditions for stability in the two universal moduli subspace was
   discussed and
   classified in \cite{Shiu:2011zt}.
   Here we focus on a model in which potentials of $H_3,\  F_0,\  F_2,\ R_6$ and O6-planes are turned on.
   The potentials we consider are given by (\ref{potentials in two moduli}).

   The stability conditions in two universal moduli subspace can be derived analytically.
   The conditions
   relate the coefficients
   of the potential and
   the
   values of the moduli at
   extrema.
   At an extremum, we can express two of the coefficients $A_{H_3}$ and $A_6$  in terms of the others and the moduli values \cite{Shiu:2011zt}
   \begin{equation}
    \begin{split}
     &A_{H_3} = - {A_{R_6} \rho^2 \over 3} + {A_{F_2} \rho^4 \over 3 \tau^2} + {A_{F_0} \rho^6 \over \tau^2}, \quad
     A_{6} = -{4 A_{R_6} \tau \over 9 \rho} - {14 A_{F_2} \rho \over 9 \tau} - {2 A_{F_0} \rho^3 \over \tau},
    \end{split}
  \end{equation}
  where $A_{H_3} >0,\ A_6 <0 $ are required.
  The stability conditions further constrain such parameters:
  \begin{equation}
   \begin{split}
    &{A_{F_2} \rho^2 \over \tau^2} < A_{R_6} < {11 A_{F_2} \rho^2 \over 7 \tau^2}, \quad
    A_{F_0} > { -7 A_{F_2}^2 \rho^4 + 9 A_{F_2} A_{R_6} \rho^2 \tau^2 - 2 A_{R_6}^2 \tau^4 \over 33 A_{F_2}\rho^6 - 21 A_{R_6} \rho^4 \tau^2}.
   \end{split}
  \end{equation}
  Applying these conditions, we can define the positive potential value at the minima by
  \begin{equation}
   V_{\rm min} = - {2 A_{F_2} \rho \over 9 \tau^4} + {2 A_{R_6} \over 9 \rho \tau^2} >0.
  \end{equation}

   \begin{figure}[t]
    \begin{center}
     \includegraphics[width=16em]{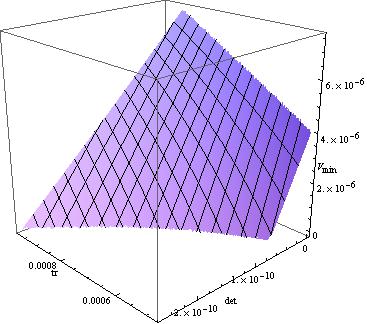}
    \end{center}
    \caption{\footnotesize Stable points at positive extrema, given values for some coefficients.
    The horizontal plane is spanned by values of the determinant and the trace of the matrix.
    The determinant is reaching zero as we increase the cosmological constant, with a fixed trace.}
    \label{fig:two-moduli}
   \end{figure}

   Following the Sylvester criteria (in linear algebra), positivity of both the determinant and the trace of a Hermitian $2\times 2$ matrix is required to ensure positivity of the entire matrix.
   In Figure \ref{fig:two-moduli}, we have a cube spanned by $V_{\rm min}$, the determinant and the trace of the matrix $\partial_{\rho_i} \partial_{\rho_j} V|_{\rm min}$.
   All points within the cube satisfy the stability conditions at a positive extremum.
   Now we see that when we fix the value of the trace, the determinant goes to zero as we increase the
   cosmological constant
   \footnote{This statement is opposite to what is observed in \cite{Marsh:2011aa}.
   This is likely because $W, F_a, Z_{ab}, U_{abc}$ are assumed to be  randomly drawn in \cite{Marsh:2011aa} whereas these quantities are related in concrete models.}.
   This corresponds to increasing the off-diagonal components to values comparable to the diagonal entries.
   Therefore the probability of a higher cosmological constant is more suppressed and less likely than that of a small cosmological constant, following the discussion in section \ref{sec:random-matrix-with}.
   Actually the upper bound of the cosmological constant is given by nothing but the
   zero
   of the determinant.
   As
   we increase the cosmological constant further in the model, we encounter a tachyon as a result of the increase in off-diagonal components.

\section{Discussions}

In this work, we use random matrix theory to estimate the probability of metastable $dS$ vacua in IIA string theory.
Our purpose here is to quantify how unlikely we can stabilize all of the K\"ahler moduli and the complex structure moduli simultaneously at a $dS$ minimum of a tree level potential. Including  quantum effects will simply yield a more complicated potential, which in turn tends to improve the randomness approximation.
Assuming the possibility of introducing some hierarchies in the mass matrix in this setup, we expect different probabilities for different situations.
Relatively large off-diagonal terms tend to make the probability ${\cal P}$ of locating a meta-stable $dS$ vacuum more Gaussianly suppressed than  exponentially suppressed.
In the IIA models, we explained why tachyons are ubiquitous in meta-stable de-Sitter extrema
 in terms of the appearance of the off-diagonal terms in the Hessian.
An increase in the cosmological constant typically
destabilizes the vacuum as a result of the increase of off-diagonal components.

Note that, in the search for meta-stable $dS$ vacua, the potential used (see (\ref{potentials in two moduli}) for example) in the search typically introduces multiple free parameters. In attempts to find meta-stable $dS$ vacua, the search allows any appropriate values for these parameters. In reality, these parameters and other hidden ones must be dynamically determined.
In the present searches, it is assumed that the modes associated with these parameters (there can be a number of modes associated with each parameter) are heavy and have been dynamically stabilized already. Furthermore, they have enough multiple stable solutions so that their values needed for $dS$ vacuum stability will be among the possible stabilized solutions. This may be a reasonable assumption for low energy scales. However, as the vacuum energy scale is increased to scales comparable to some of lighter masses of these modes, we can no longer ignore the dynamics of these modes.
For example, consider the supergravity potential (\ref{SUGRAV}) (for simplicity without the D term).
Suppose one is allowed to increase the energy scale simply by adding a positive constant to $K$, without changing the spectrum.
By doing so, we may have to include the lighter part of heavy moduli into the potential that have not been included in $K$ and $W$ so far.
The resulting relevant Hessian with increased number of moduli will not be simple in general.
So the typical relevant Hessian will grow in size and complication as the vacuum energy is increased.
Even though changing the overall scale does not alter the hierarchy between the diagonal and off-diagonal entries in the original mass matrix (which is a subset of the entire mass matrix), the increase of the number of relevant moduli increases the likelihood of instability as we argued in section \ref{sec:random-matrix-with}.
The implication of this viewpoint is as follows:
As the vacuum energy scale increases, the number of moduli to be included in the analysis of the relevant mass-squared matrix increases.
As a result, the unavoidably complicated form of the matrix will imply that the probability of having a meta-stable $dS$ vacuum becomes increasingly Gaussianly unlikely.
This reinforces our view that:
\begin{quote}
{\it The probability of finding a meta-stable de-Sitter vacuum in the string landscape becomes asymptotically Gaussianly unlikely as the vacuum energy is increased.}
\end{quote}
Since not a single classically stable Type IIA $dS$ vacuum has been found so far, this proposal/conjecture is compatible with this fact, but at the same time, we cannot check the proposal/conjecture in a non-trivial way. On the other hand, we do know some Type IIB approximate solutions; so the above proposal may be checked (to some extent) for the Type IIB regions of the landscape. However, all known meta-stable Type IIB vacua involve non-perturbative contributions to the 4-dimensional effective potential, thus the analysis is somewhat more subtle.

The probability of obtaining a $dS$ vacuum in string theory was previously
considered in \cite{Denef:2004cf}. In this seminal work, the authors estimated such probability in the context of ${\cal N}=1$ supergravity (and for simplicity, they considered cases where the D-terms are absent).
To maintain stability, the authors required the off-diagonal terms to be suppressed relative to the positive diagonal terms.
Since this potential is a function of the superpotential $W$ and its derivatives, such off-diagonal suppression is achieved if we satisfy the condition $|D_A D_B D_C W \bar{\psi}_1^A \bar{\psi}_1^B \bar{\psi}_1^C| < {\cal O}(DW)$ where the vector $\psi_1$ specifies a direction for extrema with the smallest eigenvalue, under the assumption of $|DW| \sim |W|$ (in Planck unit). Since a term proportional to $D_A D_B D_C W$ shows up in the off-diagonal components of the Hermitian mass matrix, such a constraint provides a hierarchy between the diagonal and off-diagonal entries. Now,  we have shown that the probability of stability is Gaussianly suppressed (as a function of the number of moduli) when the off-diagonal terms are within an order of magnitude of the diagonal terms; and  we typically expect one of the lightest modes to turn tachyonic first. This implies that the above condition will be Gaussianly unlikely to satisfy as the number of moduli increases.

Since the Hessian is built  from $W$ and its derivatives, the elements in the Hessian are not totally random.
However, for a complicated system with a large number of moduli, we believe that some form of the central limit theorem should  hold and our approach is valid.
While this paper was in preparation, we were informed that a more detailed model, based
on \cite{Denef:2004cf} using some random distributions of $W$ and its derivatives has recently been performed \cite{Marsh:2011aa}.

Furthermore,
in our setup, we had in mind a much wider class of constructions, some of which cannot be written  in the form of an ${\cal N}=1$ supergravity potential without $D$ terms; in these cases, our approach should provide a good estimates of the probability ${\cal P}$ when the hierarchy between the diagonal and the off-diagonal components of the Hessian can be approximately estimated.
We will report on our findings for the probability of metastable $dS$ vacua in IIB string theory in a forthcoming paper.

\section*{Acknowledgment}
We have benefited from discussions with Keith Dienes,
David Marsh, Liam McAllister,  Brooks Thomas, Thomas Van Riet,  and Timm Wrase.
GS thanks the Hong Kong Institute for Advanced Study for their hospitality.
XC is supported by the Stephen Hawking Advanced Fellowship.
GS is supported in part by a DOE grant under contract DE-FG-02-95ER40896, and a Cottrell Scholar Award from Research Corporation.
SHHT is supported in part by the National Science Foundation under grant PHY-0355005.

\appendix

\section{Other plot for random matrix\label{sec:other-plot-random}}
In this section, we show another plot for the hierarchical random matrices $M= A+B$, analyzed in section section \ref{sec:random-matrix-with}, with the fitting function ${\cal P} = a e^{-f N^g}$.

\begin{figure}[t]
 \begin{center}
  \includegraphics[width=17em]{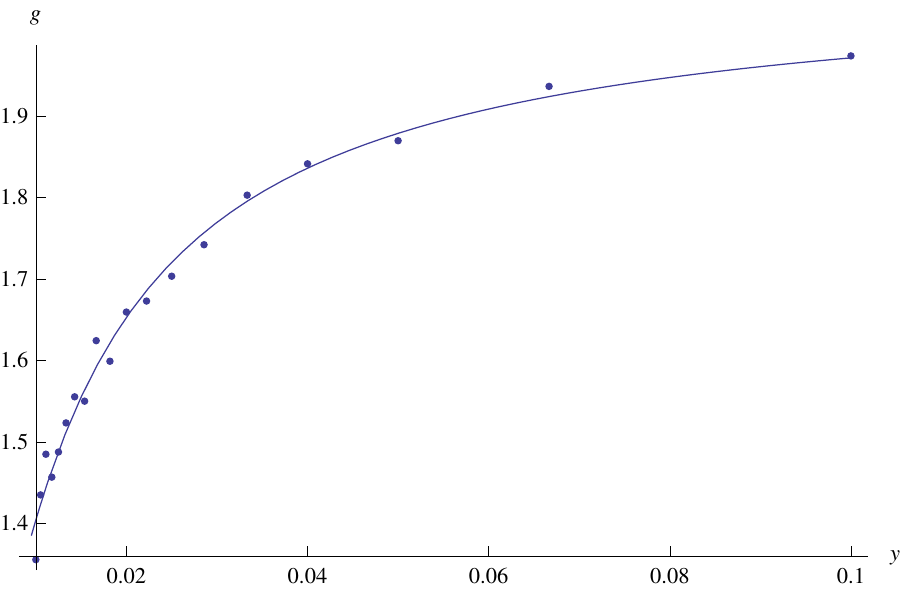}
 \end{center}
 \caption{\footnotesize Probabilities fits with the different form of probability function ${\cal P} = a e^{-f N^g}$.
 The exponent $g$ interpolates between 1 and 2 as a function $g= 1.05 + 1.03 e^{-0.0105/y}$.}
 \label{fig:coef-g}
\end{figure}

In Fig.~\ref{fig:coef-g}, we show the plot for the coefficient $g$.
The value of $g$ interpolates between 1 and 2 by a function:
\begin{equation}
 g = 1.05 + 1.03 e^{-0.0105/y}.
\end{equation}
As mentioned, this form does not have the expected large $N$ behavior with
a fixed hierarchy between diagonal and off-diagonal elements
discussed in Section \ref{sec:random-matrix-with}, especially if the fitted
value of
$g$ is not close to $2$. In these cases, it works mainly in the fitting region $N=4-20$. Once we extrapolate it to larger value of $N$, for example at $N=150,\,  \sigma_A = 100$, we get ${\cal P}_{\rm exp} = 0.157$ while the simulation suggests smaller value ${\cal P}_{\rm obs} = 0.0318$.

\bibliographystyle{utphys}
\bibliography{myrefs}

\end{document}